\documentclass[12pt]{article}
\usepackage[utf8]{inputenc}
\usepackage[margin=1in]{geometry}
\usepackage[doublespacing]{setspace}
\usepackage{graphicx} 
\usepackage{url}
\usepackage{booktabs}
\usepackage{appendix}
\usepackage{amsmath}
\usepackage{hyperref}
\usepackage{subcaption}
\usepackage{tikz}
\usepackage{tikz-cd}
\usetikzlibrary{arrows,shapes,positioning,calc,angles,quotes}
\usepackage{authblk}
\usepackage{caption}
\usepackage[backend=biber,
            style=numeric-comp,
            doi=false,
            isbn=false,
            url=false,
            eprint=false,
            sorting=none,
            maxbibnames=999]{biblatex}
\title{Large Language Models Can Be Used to Estimate the Latent Positions of Politicians}
\author[1]{Patrick Y. Wu}
\author[1,2]{Jonathan Nagler}
\author[1,2]{Joshua A. Tucker}
\author[1]{Solomon Messing}
\affil[1]{Center for Social Media and Politics, New York University}
\affil[2]{Department of Politics, New York University}
\date{Version: September 25, 2023 \\ \vspace{\baselineskip} \href{https://www.patrickywu.com/PatrickYWu_JMP1_LaMPscores.pdf}{\color{blue}Click here for the latest version}}

\begin{document}
\pagenumbering{gobble}

\begingroup
\singlespacing
\maketitle
\begin{abstract}
    \noindent Existing approaches to estimating politicians' latent positions along specific dimensions often fail when relevant data is limited. We leverage the embedded knowledge in generative large language models (LLMs) to address this challenge and measure lawmakers' positions along specific political or policy dimensions. We prompt an instruction/dialogue-tuned LLM to pairwise compare lawmakers and then scale the resulting graph using the Bradley-Terry model. We estimate novel measures of U.S. senators' positions on liberal-conservative ideology, gun control, and abortion. Our liberal-conservative scale, used to validate LLM-driven scaling, strongly correlates with existing measures and offsets interpretive gaps, suggesting LLMs synthesize relevant data from internet and digitized media rather than memorizing existing measures. Our gun control and abortion measures---the first of their kind---differ from the liberal-conservative scale in face-valid ways and predict interest group ratings and legislator votes better than ideology alone. Our findings suggest LLMs hold promise for solving complex social science measurement problems.
\end{abstract}
\endgroup

\newpage
\pagenumbering{arabic}
\setcounter{page}{1}

\section*{Introduction}
This paper outlines a novel approach to addressing challenges in measuring the latent positions of lawmakers using generative large language models (LLMs). Measuring latent positions along specific political or policy domains reduces the dimensionality of lawmakers' complex actions and stances to a low-dimensional scale. When combined with other data, these measures allow us to assess core democratic functions: how well lawmakers represent their constituents \parencite[see, e.g.,][]{ansolabehere_snyder_stewart_2001,gerber_lewis_2004,bartels_2009,thomsen_2017,caughey_warshaw_2018}, whether enacted policies have broad support or are driven by one part of the ideological spectrum \parencite[see, e.g.,][]{poole1997ideology,swers_1998,Krehbiel_1998,clinton_jackman_rivers_2004,cox_mccubbins_2005}, and how position-taking occurs outside of roll call voting \parencite[see, e.g.,][]{highton_rocca_2005,boudreau_etal_2019,russell_2021}. 

While there is broad agreement that lawmakers have positions in the space of ideology and other issue-specific dimensions, we cannot directly observe these positions---they exist in latent space and must be estimated. Existing approaches to estimating liberal-conservative ideology are based on either the behaviors or perceptions of lawmakers. Behavior-based estimates commonly use roll call votes to measure revealed preferences constrained by the legislative agenda \cite{keith_poole_nominate,poole_2005,carroll_lewis_lo_poole_rosenthal}. Other measures of behavior, such as news media sharing behavior \parencite{eady_bonneau_tucker_nagler_2020}, are also measures of liberal-conservative ideology based on revealed preferences. Liberal-conservative measures based on campaign contributions \parencite{bonica2013} assume ideological homophily in campaign giving and are based on perceptions of the contributors. Each of these measures captures a different facet of liberal-conservative ideology in a different context---interpretive gaps can occur either from modeling assumptions or a lack of relevant data.

These approaches reduce the dimensionality of a complex political space to a single left-right dimension but do not tell us about lawmakers' positions on specific issues. Specific positions on issues like gun control and abortion are difficult to measure using existing scaling approaches due to the absence of relevant data. For example, roll call votes cannot be used to measure lawmakers' stances on gun control because most sessions of Congress have no such votes on this issue. 

Generative LLMs are trained on massive corpora of internet and digitized media text, embedding much information about politics, position-taking, and widely-held perceptions as reported by journalists and other content publishers. We propose leveraging this embedded information by prompting an LLM fine-tuned to follow instructions and engage in dialogue using reinforcement learning human feedback \cite{ouyang2022training} to compare politicians on a relevant dimension. Specifically, we use ChatGPT-3.5 to pairwise compare the senators of the 116th U.S. Congress along three dimensions: liberal-conservative ideology, support of gun control support, and support of abortion rights support. We then use the Bradley-Terry Model \parencite{bradleyterry1952} to estimate a unidimensional scale measuring latent political positions of interest, which we call \textbf{La}nguage \textbf{M}odel \textbf{P}airwise comparison (LaMP) scores. 

\begin{figure}[!ht]
\centering
\begin{tikzpicture}[scale=0.75,every node/.style={scale=0.75}]
\node[draw, shape=rectangle, minimum width=100pt, minimum height=50pt, align=center] (step1) {Create matchups\\between items\\to be scaled};
\node[draw, shape=rectangle, minimum width=100pt, minimum height=50pt, right=10pt of step1, align=center] (step2) {Create\\comparison\\prompts};
\node[draw, shape=rectangle, minimum width=100pt, minimum height=50pt, right=10pt of step2, align=center] (step3) {Input comparison\\prompts into\\generative LLM};
\node[draw, shape=rectangle, minimum width=100pt, minimum height=50pt, right=10pt of step3, align=center] (step4) {Extract answers\\(identify ``winner''\\and ``loser'' of\\each matchup)};
\node[draw, shape=rectangle, minimum width=100pt, minimum height=50pt, right=10pt of step4, align=center] (step5) {Use the\\Bradley-Terry\\model to scale\\the extracted answers};

\draw[->] (step1) -- (step2);
\draw[->] (step2) -- (step3);
\draw[->] (step3) -- (step4);
\draw[->] (step4) -- (step5);
\end{tikzpicture}
\caption{An overview of the proposed pairwise comparison approach with instruction/dialogue-tuned generative LLMs.}
\label{fig:overview}
\end{figure}
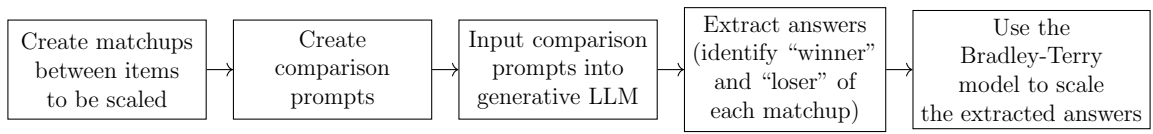

These pairwise comparisons are made in a zero-shot learning setting because we do not provide ChatGPT with \textit{any additional information} besides the senator's name, party affiliation, and state represented, and we \textit{do not} include any examples of pairwise comparisons. For the liberal-conservative ideology scale, we prompt ChatGPT to pick the senator who is more conservative (or liberal) for a given pair of senators. For the gun control scale, we prompt ChatGPT to pick the senator more likely to support gun control. For the abortion rights scale, we prompt ChatGPT to pick the senator more likely to be pro-choice (or pro-life). Figure \ref{fig:overview} shows an overview of the proposed pairwise comparison approach. 

The liberal-conservative ideology scale has been extensively studied in the U.S. national legislature, providing a widely accepted and well-validated set of measures by which we can validate LLM-driven scaling and better understand its strengths. Our Ideology LaMP scores highly correlate with the first dimension of DW-NOMINATE, the most popular liberal-conservative scaling of senators \parencite{keith_poole_nominate,poole_2005,carroll_lewis_lo_poole_rosenthal}. However, Ideology LaMP scores also depart from DW-NOMINATE scores in important ways. For example, our approach places senators who vote \textit{against} their party for ideologically extreme reasons on the ends, while DW-NOMINATE places them more toward the center. We also find that Ideology LaMP scores predict human evaluations of the ideologies of senators better than other measures \parencite{hopkins_noel_2022}, including DW-NOMINATE. 

The gun control scale has not been estimated in the political science literature because of a lack of data on the behaviors and perceptions of the senators concerning gun control. We find that our Gun Control LaMP scale places senators in sensible positions. It does not simply imitate Ideology LaMP scores. For example, Bernie Sanders, the most liberal senator on the Ideology LaMP scale, is placed in the middle of the Democratic senators on the Gun Control LaMP scale. Validating this measure, we find that Gun Control LaMP scores predict the grades given to each senator by the National Rifle Association (NRA) and individual votes on the 2022 Bipartisan Safer Communities Act, the latter of which is out-of-sample---the vote took place after ChatGPT-3.5's training data ends.

Similarly, abortion rights scales have not been estimated in the political science literature for reasons related to absent or sparse data. Our Abortion Rights LaMP scores place senators in sensible positions, such as placing self-described pro-choice Republicans Susan Collins and Lisa Murkowski among the Democratic senators. The Abortion Rights LaMP scores also predict the grades given to each senator by NARAL Pro-Choice America better than DW-NOMINATE. These issue-specific scales offer advantages over using interest group ratings: they eliminate the issue of ``lumpiness,'' such as all Democratic senators receiving `Fs' from the NRA, and provide a single measure of all senators, avoiding the need to combine interest group ratings that are based on different criteria \parencite{mccarty_measuring_legislative_preferences}.

Partial correlations analysis suggests that LaMP scores are a blend of textually available information about individual lawmakers. In other words, LaMP scores reflect both behaviors, such as votes, floor speeches, and commentary about other lawmakers, as well as perceptions of these lawmakers, such as news stories, blog posts, and editorials. Because of the black box nature of LLMs, we do not know how the LLM weighs the underlying textual information given a prompt. At the same time, the scope of the LLM's embedded knowledge base underpinning the measures is its strength, making up for interpretive gaps in existing measures of ideology and allowing us to measure politicians' stances along specific issues.

We cannot simply prompt an LLM to return a list of lawmakers ranked along one of these dimensions for several reasons. First, ChatGPT, in particular, often does not consistently return a complete list of senators ranked from most liberal to most conservative: it will sometimes only return partial lists, generate inconsistent lists across repeated promptings, or refuse to create such lists. Pairwise comparisons also enable us to establish a scale where the differences are meaningful, whereas ordinal rankings indicate order without quantifying the gaps between ranks. For this reason, researchers have used pairwise comparisons extensively in social science scaling applications. They are also easier to complete. For example, \textcite{LOEWEN2012212} conduct a survey experiment with pairwise comparisons and the Bradley-Terry model to determine the most persuasive arguments. \textcite{carlson_montgomery_2017}'s SentimentIt R package conducts pairwise comparisons to label political texts. \textcite{hopkins_noel_2022}, a closely related work to this study, use pairwise comparisons among political activists to scale senators of the 114th Congress and the 117th Congress along the liberal-conservative continuum.

In summary, we find that when prompted with pairwise comparisons, the LLM does not hallucinate; it does not simply parrot pre-existing measures of the ideologies of senators; and, most importantly, we can prompt it to evaluate pairwise comparisons that analysts can use to construct novel scales that would be impossible to do with existing scaling methods. Our validation and analysis of Ideology LaMP scores suggest that we can use a class of instruction/dialogue-tuned LLMs to create a scale that correlates with an amalgamation of information about policy positions, voting behaviors, campaign-giving patterns, and public perceptions about politicians. The validity of Gun Control LaMP scores and Abortion Rights LaMP scores suggest that complex issue-specific scales can be estimated using pairwise comparison scaling with generative LLMs. Overall, our proposed pairwise scaling approach can give us a more comprehensive understanding of legislative behaviors and policy preferences.

\section*{Results}
We use ChatGPT to make pairwise comparisons about liberal-conservative ideology, gun control, and abortion, and then scale its answers using the Bradley-Terry model, producing LaMP scores. We use the liberal-conservative ideology scale to better understand the strengths of LaMP scores. The gun control scale and abortion rights scale demonstrate how we can use the approach to estimate novel issue-specific scales. We call the pairwise comparisons ``matchups.'' There are 5,151 total matchups across all senators in the 116th Congress for each scale.

\subsection*{Liberal-Conservative Ideology Scaling}
For liberal-conservative ideology scaling, the ``winner'' of each matchup was the senator ChatGPT answered as being more conservative in its response. ``Winners'' are assigned this way to intuitively place more conservative senators on the right side of the scale and more liberal senators on the left side of the scale. We call the resulting scores ``Ideology LaMP scores.'' We highlight interesting features of Ideology LaMP scores and analyze their relationship with pre-existing measures of these senators' ideologies within and across parties. The Methods section details the prompts used for these pairwise comparisons. 

\subsubsection*{Ideology LaMP scores highly correlate across repeated iterations}
We ran the entire set of matchups across all senators three times. We look at the correlation of the Ideology LaMP scores generated by each set of complete matchups. Among the three iterations, the lowest correlation between any two iterations' LaMP scores was 0.997. ChatGPT's responses to pairwise comparisons are highly consistent in repeat interations. Given the high correlations, we use Ideology LaMP scores estimated using all matchups across all iterations for the rest of the analysis. 

\subsubsection*{Ideology LaMP scores highly correlate with DW-NOMINATE} 
\begin{figure}[!ht]
    \centering
    \includegraphics[width=\textwidth]{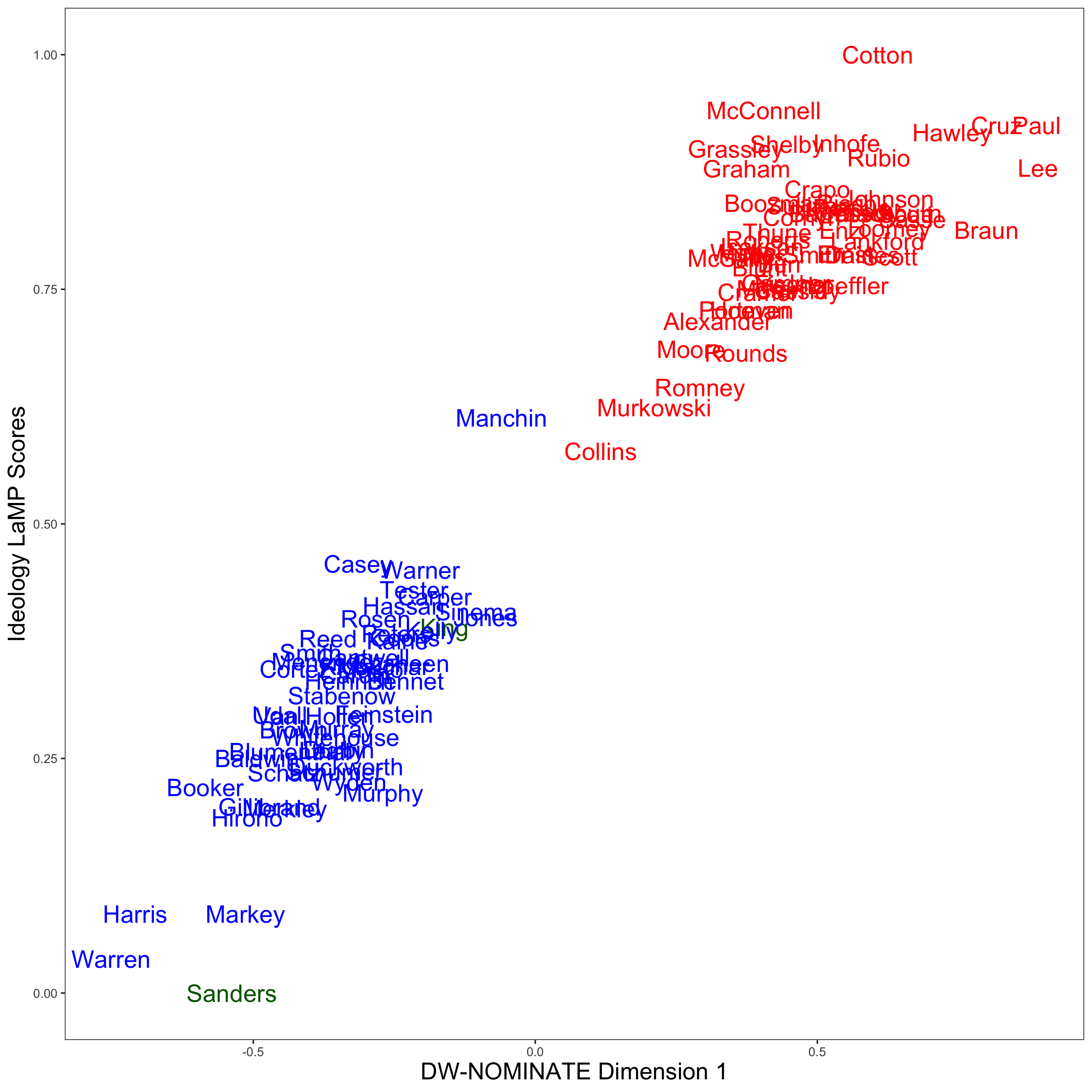}
    \caption{First Dimension of DW-NOMINATE vs. Ideology LaMP scores. Democratic senators are in blue, Republican senators are in red, and Independent senators are in green.}
    \label{fig:nominate_v_LaMPscores}
\end{figure}

DW-NOMINATE (Dynamic, Weighted NOMINAl Three-Step Estimation) is a multidimensional scaling approach that uses roll call voting patterns to estimate the ideological positions of legislators \parencite{keith_poole_nominate,poole1997ideology,poole_2005,carroll_lewis_lo_poole_rosenthal}. It is the most widely used measure of legislator ideology \parencite{caughey_schickler_2016}. The first dimension of DW-NOMINATE is typically interpreted as the liberal-conservative continuum in United States politics \parencite{poole1997ideology}. The overall correlation between Ideology LaMP scores and the first dimension of DW-NOMINATE is 0.967. They correlate at 0.838 among Democratic senators and correlate at 0.649 among Republican senators. Figure \ref{fig:nominate_v_LaMPscores} compares the first dimension of DW-NOMINATE against Ideology LaMP scores.

\subsubsection*{Ideology LaMP scores do not simply parrot DW-NOMINATE}
Notably, our method estimates Joe Manchin to be more conservative than Susan Collins. Such a placement intuitively makes sense: for example, on the issue of abortion, Manchin is pro-life, while Collins is pro-choice. In contrast, there is no overlap between senators of opposing parties in the first dimension of DW-NOMINATE.

Looking at the extremes also indicates that ChatGPT is not simply recalling DW-NOMINATE scores for senators when responding to pairwise assessments. DW-NOMINATE places Elizabeth Warren and Kamala Harris as the most liberal senators. Ideology LaMP scores, on the other hand, place Bernie Sanders and Elizabeth Warren as the most liberal senators. This comports with surveyed political activists, who also named Sanders and Warren the most liberal senators \parencite{hopkins_noel_2022}. Sanders' placement towards the center of the first dimension of DW-NOMINATE is the result of Sanders occasionally voting against the Democratic party \parencite{duck-mayr_montgomery_2023}. LaMP scores, conversely, likely pick up signals from not only roll call votes but also sources such as, among other things, mainstream news articles that discuss Sanders' left-leaning positions on various issues. Among Republicans, Ideology LaMP scores rank Tom Cotton and Mitch McConnell as the most conservative senators; DW-NOMINATE ranks Mike Lee and Rand Paul as the most conservative. Surveyed political activists ranked Cruz and Cotton as the most conservative senator \parencite{hopkins_noel_2022}\footnote{Sessions was ranked the second most conservative senator among political activists, but he was not in the 116th Senate.}; Cruz was the third most conservative senator according to Ideology LaMP scores.  

Comparing the ordinal rankings of DW-NOMINATE and Ideology LaMP scores, senators differed, on average, by 8.31 positions. Some of the largest differences in ordinal rankings were Chuck Grassley (DW-NOMINATE: 48th most conservative; LaMP scores: 8th most conservative), Mitch McConnell (DW-NOMINATE: 38th most conservative; LaMP scores: 2nd most conservative), and Lindsey Graham (DW-NOMINATE: 45th most conservative; LaMP scores: 11th most conservative). Among the 10 senators with the largest differences in ordinal rankings, 9 are Republicans. These differences seem to be shaped by their public stances with respect to Donald Trump, his policies, and his nominees, such as Chuck Grassley's vocal support for Brett Kavanaugh's nomination to the Supreme Court of the United States. Again, DW-NOMINATE would not capture these public stances.

\subsubsection*{LaMP scores highly correlate with alternative measures of ideology} 
Next, we compare Ideology LaMP scores to two alternative measures of ideology from the political science literature that are not based on roll call votes but on the perceptions of the senators: one scales senators based on political activists' knowledge, and the other on patterns in campaign donations. Specifically, we look at the two following measures:
\begin{enumerate}
    \item Perceived ideology scores \parencite{hopkins_noel_2022}: Perceived ideology scores are estimated using political activists' answers to pairwise comparisons of senators. Perceived ideology scores can be considered a separate but related measure of ideology: they capture the perceived ideological positions of politicians, which shape how people vote and interact with politicians. Perceived ideologies can differ from how politicians view themselves ideologically and what their revealed preferences are. \textcite{hopkins_noel_2022} included these pairwise comparisons of senators of the 117th Congress in a YouGov survey in April 2021. The authors had 1,110 activists answer these pairwise comparisons; they then scaled the activists' answers using the Bradley-Terry model. The 11 senators who retired or did not secure a new term at the end of the 116th Congress were not included in their survey. These perceived ideology scores offer a way to compare a scale estimated using ChatGPT's pairwise comparisons with a scale estimated using human-labeled pairwise comparisons. 
    \item Campaign Finance Scores \parencite{bonica2013}: \textbf{C}ampaign \textbf{F}inance scores (CFscores) are a measure of the ideologies of politicians, donors, and interest groups. CFscores are estimated using a network that links all individual contributors to all political candidates who received donations. It assumes that individuals choose to give to candidates close to them in a latent ideological space and scales all actors in that space; in other words, it measures ideology based on the donors' perceptions of lawmakers. We used each senator's latest CFscore; the Database on Ideology, Money in Politics, and Elections have estimated CFscores up to the 2018 election cycle \parencite{bonica_dime}. We look at the recipient CFScore, which is the estimated ideology of the senator based on donations received. Tammy Baldwin, Mark Kelly, and Kelly Loeffler are missing recipient CFscores.
\end{enumerate}

\begin{figure}[!ht]
    \centering
    \begin{subfigure}[b]{0.495\textwidth}
        \includegraphics[width=\textwidth]{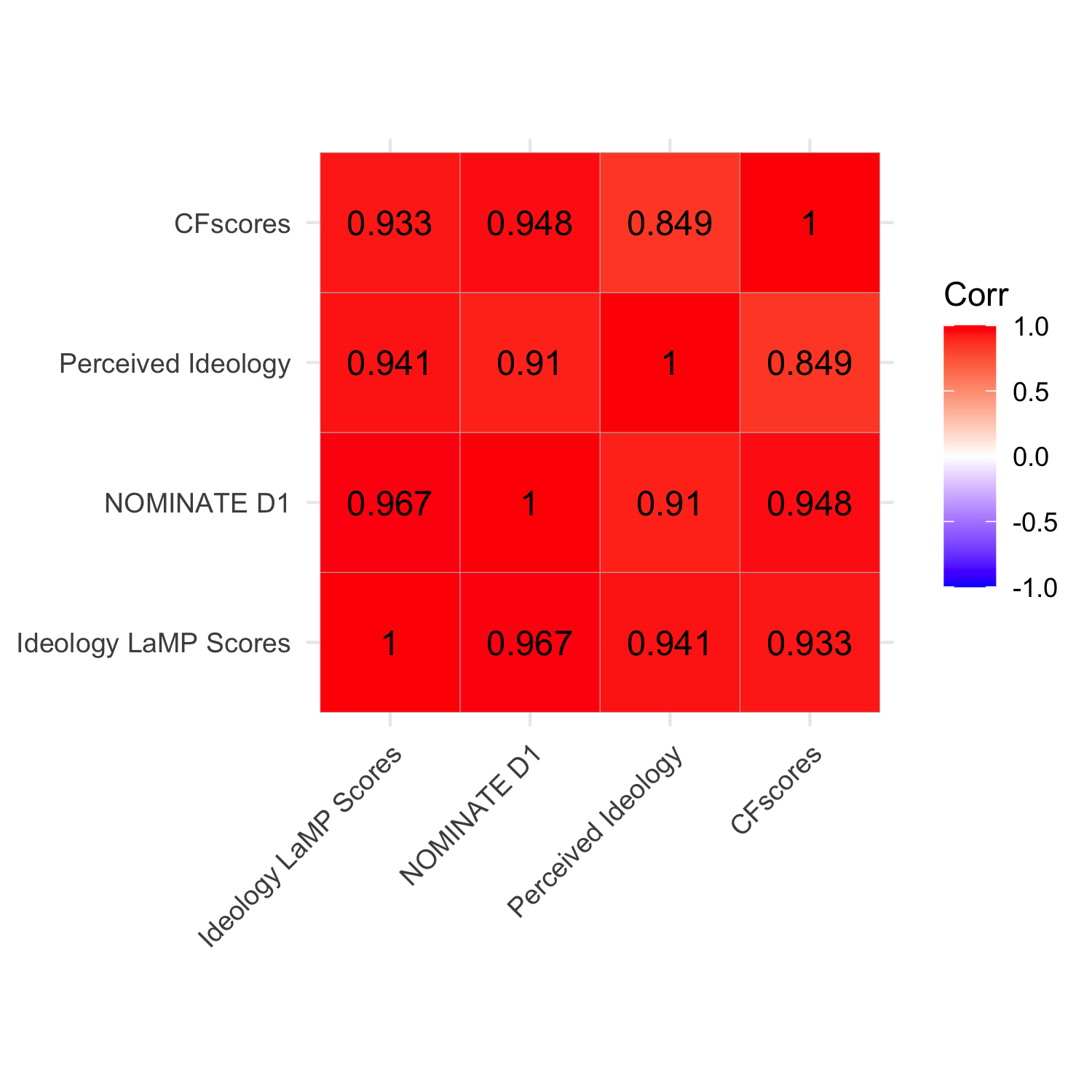}
        \caption{All Senators}
        \label{fig:overall_corrs_subplot}
    \end{subfigure}
    \begin{subfigure}[b]{0.495\textwidth}
        \includegraphics[width=\textwidth]{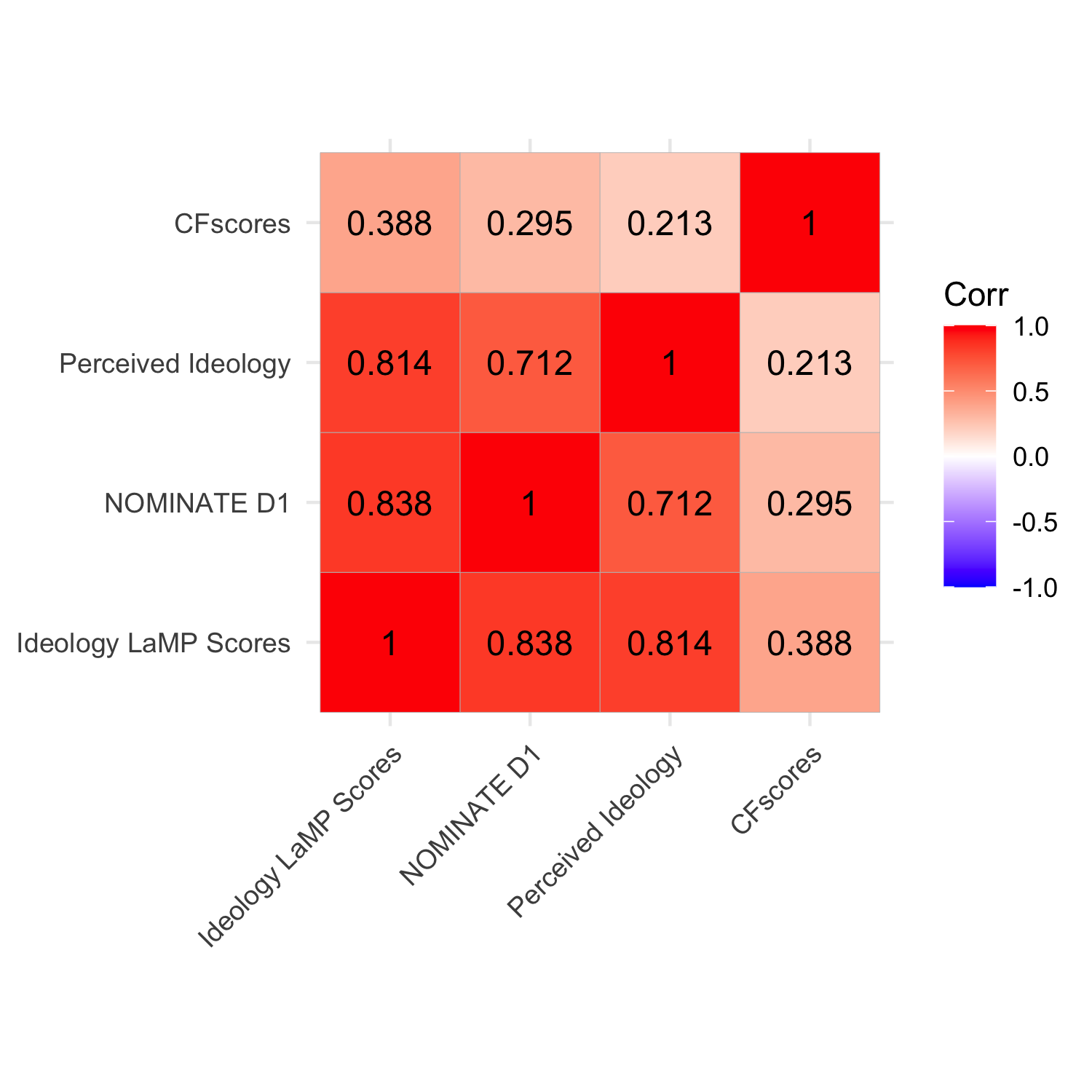}
        \caption{Democratic Senators}
        \label{fig:dems_corrs_subplot}
    \end{subfigure}
    \begin{subfigure}[b]{0.495\textwidth}
        \includegraphics[width=\textwidth]{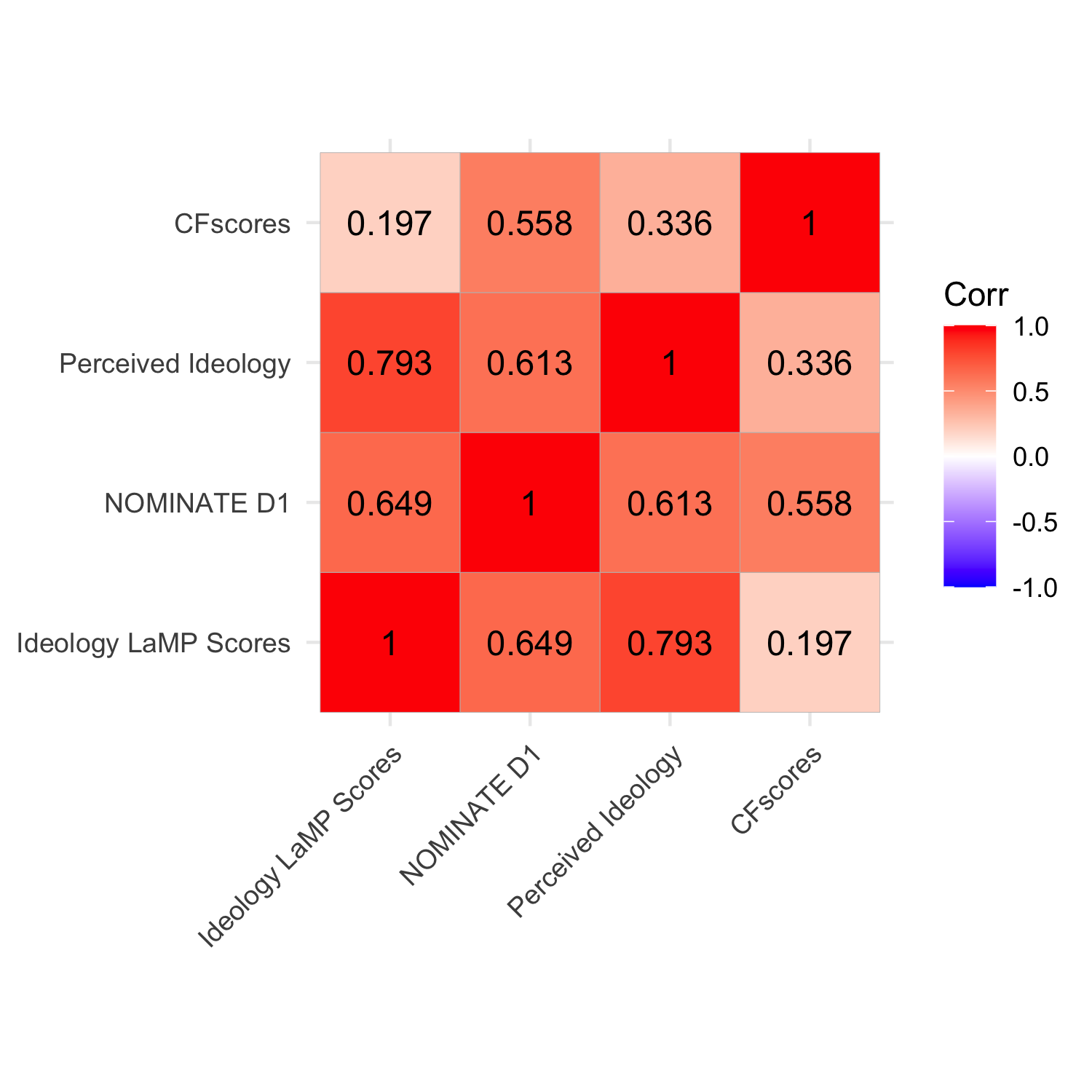}
        \caption{Republican Senators}
        \label{fig:reps_corrs_subplot}
    \end{subfigure}
    \caption{Correlation matrices of LaMP scores, the first dimension of DW-NOMINATE, \textcite{hopkins_noel_2022}'s perceived ideology scores, and \textcite{bonica2013}'s CFscores.}
    \label{fig:correlation_plots}
\end{figure}

Figure \ref{fig:correlation_plots} shows bivariate analyses across Ideology LaMP scores, DW-NOMINATE Dimension 1, perceived ideology scores, and CFscores. Across all senators, Ideology LaMP scores highly correlate with perceived ideology (0.941) and CFscores (0.933). 

\subsubsection*{LaMP scores reflect both behaviors and perceptions}
Table \ref{tab:partial_corrs} shows the partial correlations between Ideology LaMP scores and the three measures of ideology as described above. DW-NOMINATE is a measure of ideology based on the behaviors of senators, while perceived ideology and CFscores are measures of ideology based on perceptions of the senators. For each cell, the partial correlation between the Ideology LaMP scores and the measure in the column title is calculated controlling for the other two measures of ideology. The partial correlations suggest that no single measure of ideology fully explains Ideology LaMP scores. Instead, the results indicate that Ideology LaMP scores reflect a measure of ideology based on both behaviors and perceptions of the senators. This interpretation holds when we look at the partial correlations across all senators, Democratic senators, and Republican senators, except for the partial correlation between Ideology LaMP scores and CFscores among Republican senators. To be clear, it does not suggest that these measures collectively capture everything related to behaviors and perceptions about the senators. However, these three measures of ideology capture a wide range of revealed preferences and perceptions of the senators. 

\begin{table}[!ht]
\centering
\begin{tabular}{l|ccc}
                                    & DW-NOMINATE         & Perceived Ideology      & CFscores       \\ \hline
Ideology LaMP Scores, All Senators  & $0.441^{***}$       & $0.621^{***}$           & $0.300^{**}$   \\
Ideology LaMP Scores, Dem. Senators & $0.578^{***}$       & $0.592^{***}$            & $0.334^{*}$    \\
Ideology LaMP Scores, Rep. Senators & $0.467^{**}$        & $0.678^{***}$           & $-0.360^{*}$   \\ \hline
\multicolumn{4}{l}{\scriptsize{Note: $^{***}p<0.001$; $^{**}p<0.01$; $^{*}p<0.05$}}
\end{tabular}
\caption{Partial correlations between Ideology LaMP scores and the first dimension of DW-NOMINATE, perceived ideology, and CFscores. Each cell shows the partial correlations between Ideology LaMP scores and the measure in the column title, controlling for the other two measures of ideology. P-values are calculated using the t-statistic described in \textcite{kim2015ppcor}.} 
\label{tab:partial_corrs}
\end{table}

\subsubsection*{LaMP scores better predict human evaluations of ideology}
\begin{figure}[!ht]
    \centering
    \includegraphics[width=\textwidth]{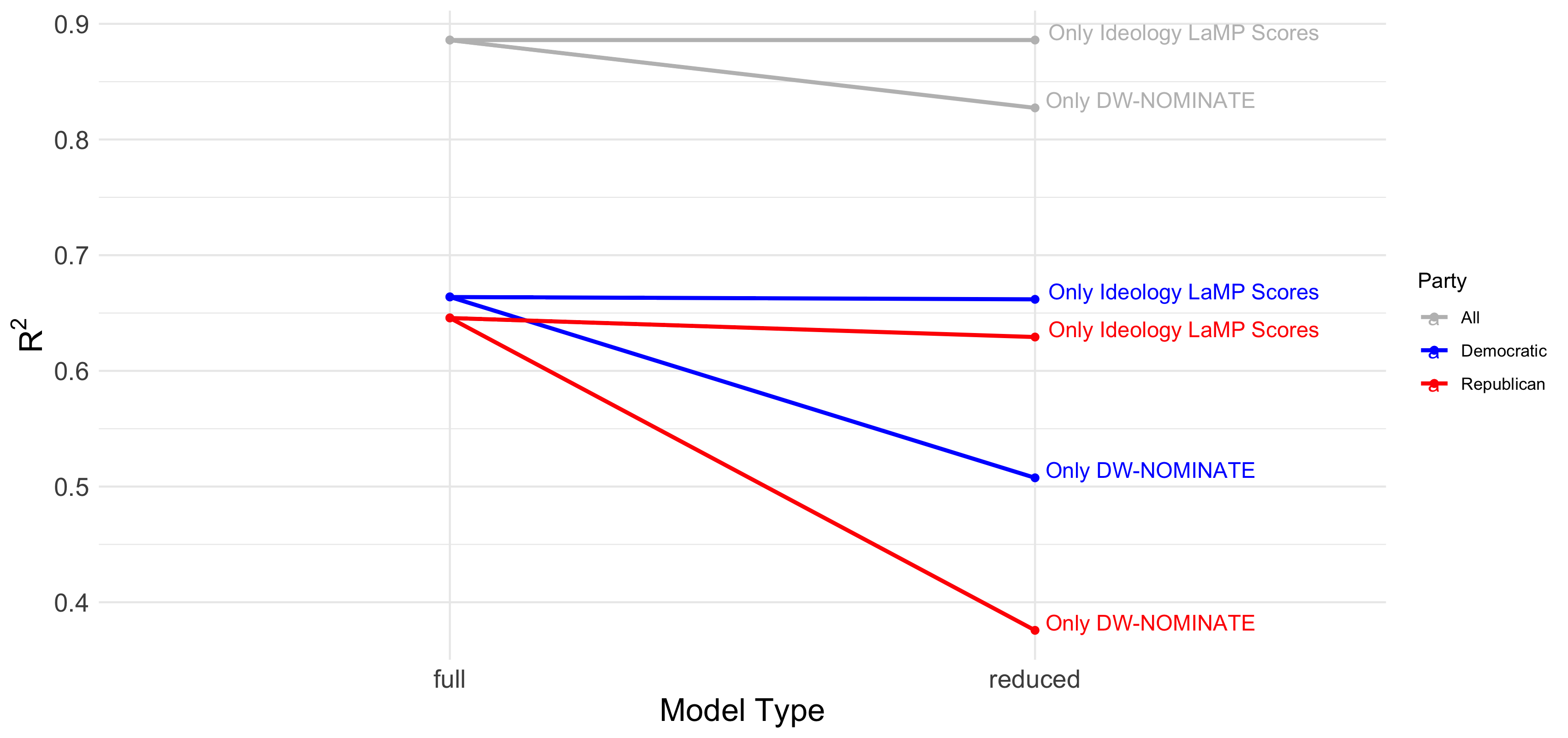}
    \caption{Comparing the proportion of variance, $R^2$, between the full and reduced models. The full model regresses perceived ideology scores on both the first dimension of DW-NOMINATE and Ideology LaMP scores. The reduced models only use DW-NOMINATE (denoted as ``Only DW-NOMINATE'') or Ideology LaMP scores (denoted as ``Only Ideology LaMP scores'') as predictors. The proportion of variance explained, $R^2$, is always lower when Ideology LaMP scores are removed as a predictor than when DW-NOMINATE is removed.}
    \label{fig:pi_predicted_using_nominate_cgs}
\end{figure}

Figure \ref{fig:correlation_plots} indicates that the correlation between Ideology LaMP scores and perceived ideology scores is higher than the correlations between perceived ideology and the other measures of ideology when looking at all senators and senators by party. To more formally analyze the predictive power of Ideology LaMP scores on human evaluations of senators' ideologies, we compare the predictive power of Ideology LaMP scores and the first dimension of DW-NOMINATE, the second-highest correlating measure with perceived ideology scores. We use multivariate analyses to calculate how much the proportion of variance explained ($R^2$) in perceived ideology scores falls when we compare the full model, regressing perceived ideology scores on both the first dimension of DW-NOMINATE and Ideology LaMP scores, with reduced models, which only use Ideology LaMP scores or DW-NOMINATE as the predictor. Figure \ref{fig:pi_predicted_using_nominate_cgs} shows how the proportion of variance explained in perceived ideology changes as we move from the full to reduced models. 

The proportion of variance explained in perceived ideology always falls more when we drop Ideology LaMP scores as a predictor. We statistically confirm these results using partial F-tests. When looking across all senators, only Democratic senators, and only Republican senators, the partial F-test p-values for the full model and the reduced model with only DW-NOMINATE as a predictor is significant ($p<.0001$ for all F-test p-values), while the p-value for the full model and the reduced model with only Ideology LaMP scores as a predictor is not significant ($p>.16$ for all F-test p-values). It suggests that ChatGPT uses synthesized information that is highly correlated with how activists perceive these senators' ideologies. It also further indicates that Ideology LaMP scores go beyond measuring position-taking.

\subsection*{Gun Control Support Scaling}
To scale senators along a gun control support continuum, we prompt the LLM to judge which senator is more likely to support gun control. We call the resulting scores ``Gun Control LaMP scores.'' We highlight features of Gun Control LaMP scores and analyze their relationship with interest group ratings and voting behaviors. 

\subsubsection*{Gun Control LaMP scores highly correlate across repeated iterations}
We ran the entire set of matchups across all senators three times. The lowest correlation among any two iterations was 0.993. Again, ChatGPT's responses to pairwise comparisons about gun control support are consistent despite this being a specific area of public policy. 

\subsubsection*{Gun Control LaMP scores differ from Ideology LaMP scores}
\begin{figure}[!ht]
    \centering
    \includegraphics[width=\textwidth]{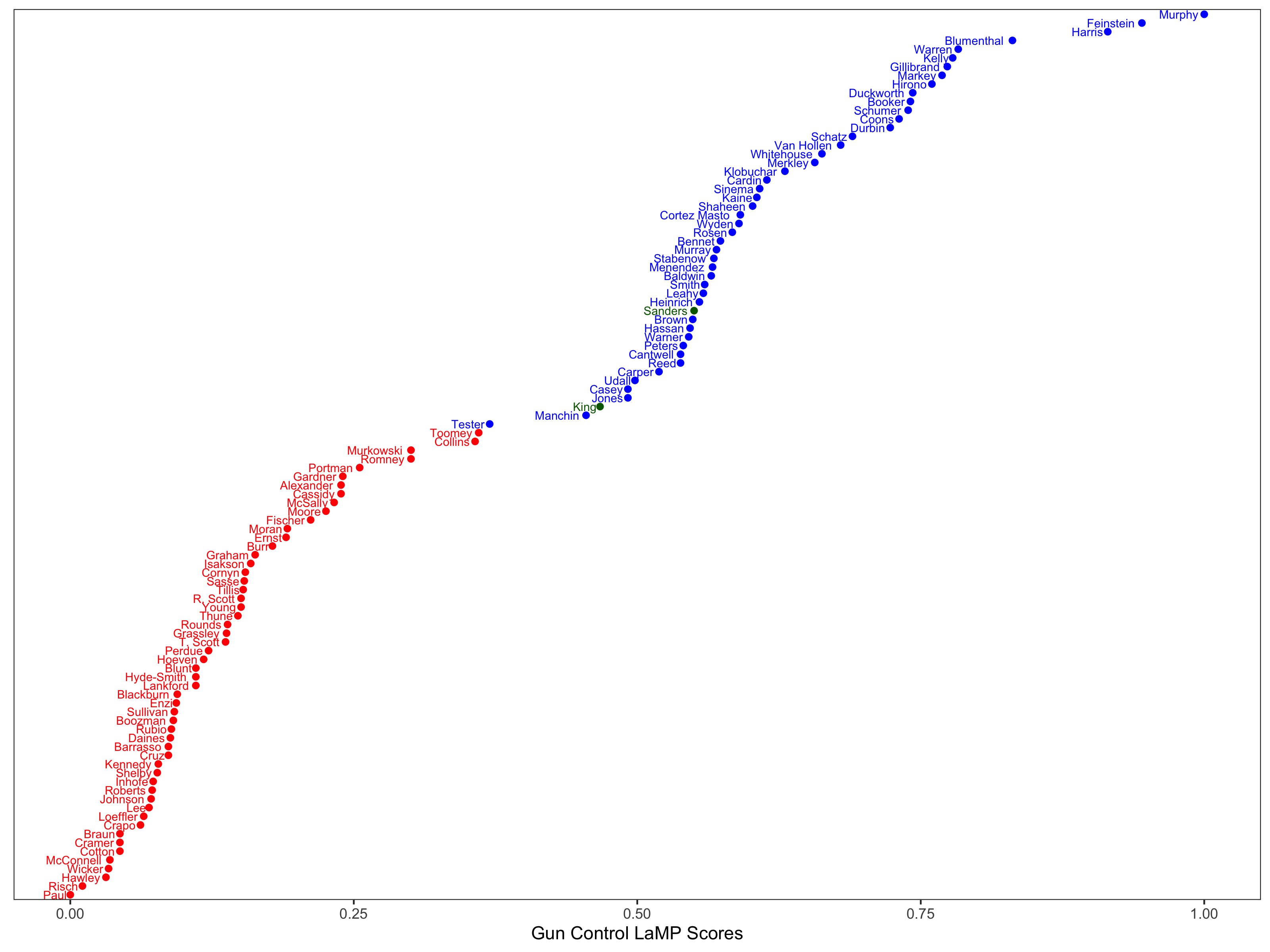}
    \caption{Gun Control LaMP scores across all senators. Democratic senators are in blue, Republican senators are in red, and Independent senators are in green.}
    \label{fig:overall_gcs_LaMPscores}
\end{figure}

The Gun Control LaMP scores of all senators of the 116th Congress are illustrated in Figure \ref{fig:overall_gcs_LaMPscores}. This issue-specific scale differs from Ideology LaMP scores. We note that we expect negative correlations because of party alignment on the issue of gun control. 

Overall, they correlate -0.943, but this correlation is largely driven by party alignment on the issue of gun control. Within each party, the correlations are not as strong: among Republican senators, the correlation is -0.684, and among Democratic senators (including the two independent senators), the correlation is -0.570. We also note that there is no overlap between the senators of the two parties, unlike the Ideology LaMP scores. 

Comparing Gun Control LaMP scores with Ideology LaMP scores, we find face validity with the Gun Control LaMP scores. For example, Ideology LaMP scores place Mark Kelly as a centrist Democratic senator, but Gun Control LaMP scores place him as one of the strongest gun control supporters. This aligns with his outspoken advocacy for gun control following the attempted assassination of his wife, former U.S. representative Gabby Giffords. On the other hand, Bernie Sanders, the most liberal Democratic senator based on Ideology LaMP scores, is placed in the middle among the Democratic senators on this issue-specific scale. Bernie Sanders often treads carefully on the issue of gun control, reflecting his support of the hunting traditions of his home state of Vermont. Pat Toomey, placed in the middle of the Republican Party on the overall liberal-conservative ideology scale, is placed as the Republican most supportive of gun control. Toomey, breaking from his party, has supported background checks and state red flag laws. 

\subsubsection*{Gun Control LaMP scores better predict NRA grades}
\begin{figure}
    \centering
    \includegraphics[width=\textwidth]{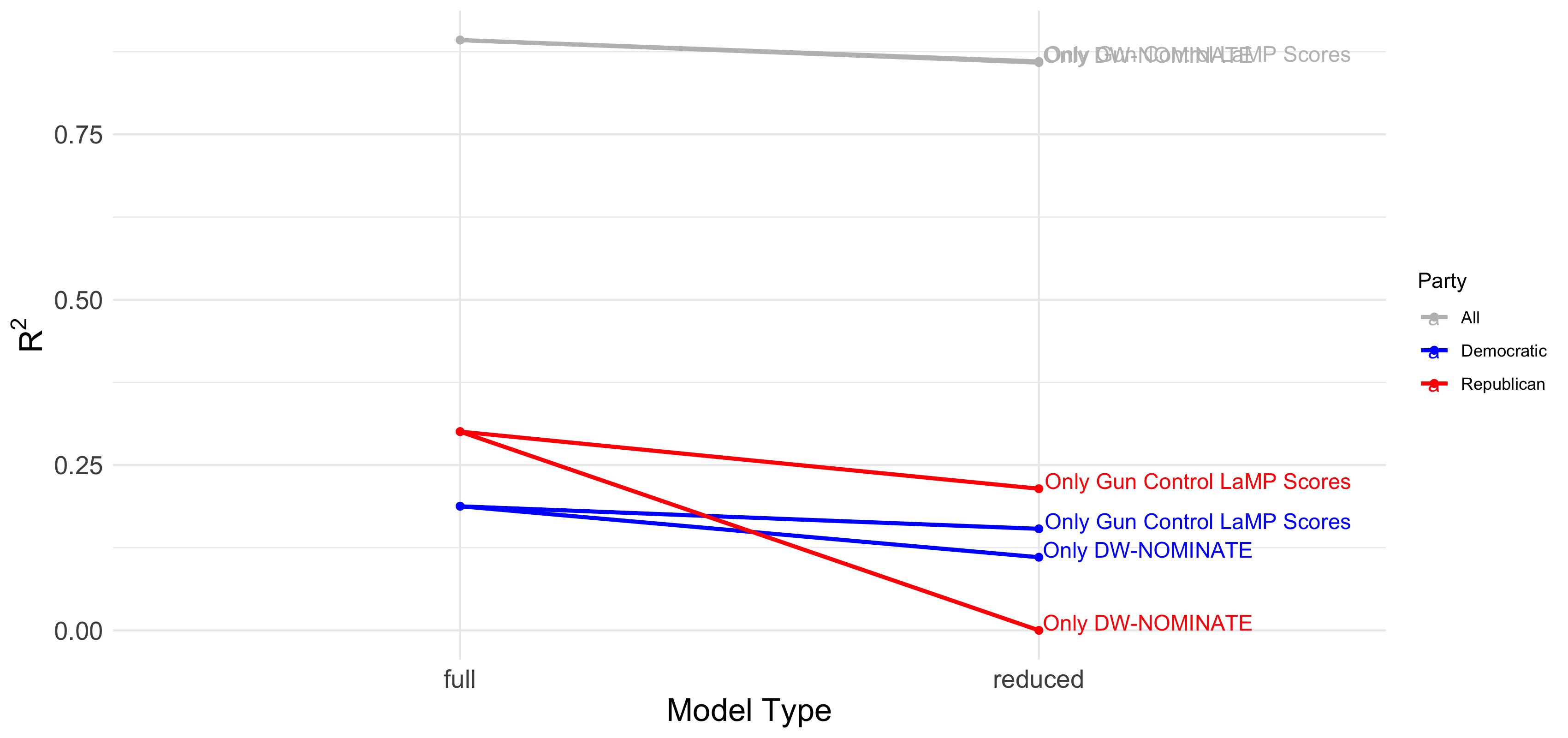}
    \caption{Comparing the proportion of variance, $R^2$, between the full and reduced models. The full model regresses the NRA grades on both the first dimension of DW-NOMINATE and the Gun Control LaMP scores. The reduced models only use DW-NOMINATE or Gun Control LaMP scores as predictors. The $R^2$ is always lower when Gun Control LaMP scores are removed as a predictor than when DW-NOMINATE is removed.}
    \label{fig:nra_predictive}
\end{figure}

We compare the predictive power of Gun Control LaMP scores and the first dimension of DW-NOMINATE on NRA grades. The NRA assigns grades each election cycle, although not all candidates receive one. We collected the latest available NRA grade for each senator from Vote Smart. We again use multivariate analyses to calculate how much the proportion of variance explained ($R^2$) in NRA grades falls when we compare the full model, regressing the NRA grade on both the first dimension of DW-NOMINATE and Gun Control LaMP scores, with reduced models, which only use DW-NOMINATE or the Gun Control LaMP scores as the predictor. Figure \ref{fig:nra_predictive} shows how $R^2$ falls as we move from full to reduced models across all senators, across the Democratic senators, and across the Republican senators. 

The proportion of variance explained in NRA grades always falls more when we drop Gun Control LaMP scores as a predictor than when dropping DW-NOMINATE as a predictor. Looking at the Republican senators in particular, the $R^2$ is near 0 when only using DW-NOMINATE to predict the NRA grades of Republicans. However, the $R^2$ is 0.214 when only using the Gun Control LaMP scores, indicating that the Gun Control LaMP scores measure a different latent construct than DW-NOMINATE. The same pattern holds for Democratic senators, although the difference is smaller. Because the NRA treats Democratic senators differently from Republican senators and gives nearly all of them scores close to 0---Tom Udall had the highest grade from the NRA among Democratic senators, at 21---the $R^2$ among Democratic senators is less interpretable because there is so little variance to be explained. Among all senators, the difference in $R^2$ is negligible, largely because positions on gun control are still aligned with the two parties.

\subsubsection*{Gun Control LaMP scores predict Republican votes on the 2022 Bipartisan Safer Communities Act}
ChatGPT's training data could explain the predictive power of Gun Control LaMP scores on NRA grades. To evaluate the scale's external validity, we predict Republican votes on the 2022 Bipartisan Safer Communities Act, which could not have been used in ChatGPT's training data because the model was only trained on data through September 2021. 15 Republican senators voted alongside all Democratic senators to pass the bill; thus, we exclude the votes of Democratic senators. We used a logistic regression predicting Republican votes on the bill using Gun Control LaMP scores and the first dimension of DW-NOMINATE; results are in Table \ref{tab:bipartisan_votes_gcscs}. Gun Control LaMP scores are a statistically significant predictor of Republican votes on the 2022 Bipartisan Safer Communities Act, even when controlling for the first dimension of DW-NOMINATE. 

\begin{table}[!ht] \centering 
\begin{tabular}{@{\extracolsep{5pt}}lc} 
\\[-1.8ex]\hline 
\hline \\[-1.8ex] 
\\[-1.8ex] & Voted Yea \\ 
\hline \\[-1.8ex] 
 Gun Control LaMP scores & 21.484$^{**}$ \\ 
  & (8.258) \\ 
  & \\ 
 DW-NOMINATE & $-$9.863 \\ 
  & (5.447) \\ 
  & \\ 
 Constant & 0.882 \\ 
  & (2.745) \\ 
  & \\ 
\hline \\[-1.8ex] 
Observations & 45 \\ 
Log Likelihood & $-$15.202 \\ 
\hline 
\hline \\[-1.8ex] 
\textit{Note:}  & \multicolumn{1}{r}{$^{*}$p$<$0.05; $^{**}$p$<$0.01; $^{***}$p$<$0.001} \\ 
\end{tabular}
\caption{We used logistic regression to predict Republican votes on the 2022 Bipartisan Safer Communities Act using the Gun Control LaMP scores and the first dimension of DW-NOMINATE as predictor variables. Gun Control LaMP scores are statistically significantly predictive of Republican votes on this bill. Democratic votes were excluded because all Democratic senators voted in favor of the bill.} 
\label{tab:bipartisan_votes_gcscs} 
\end{table}

\subsection*{Abortion Rights Support Scaling}
To scale senators along an abortion rights support continuum, the ``winner'' of each matchup was the senator the LLM answered as being more pro-choice (or more pro-life). We call these ``Abortion Rights LaMP scores.''

\subsubsection*{Abortion Rights LaMP scores highly correlate across repeated iterations}
Similar to the other applications, we find that the lowest correlation among any two iterations was 0.996. It again demonstrates ChatGPT's consistency in its responses to pairwise comparisons despite being asked about a specific policy area. 

\subsubsection*{Abortion Rights LaMP scores have high face-validity}
\begin{figure}[!ht]
    \centering
    \includegraphics[width=\textwidth]{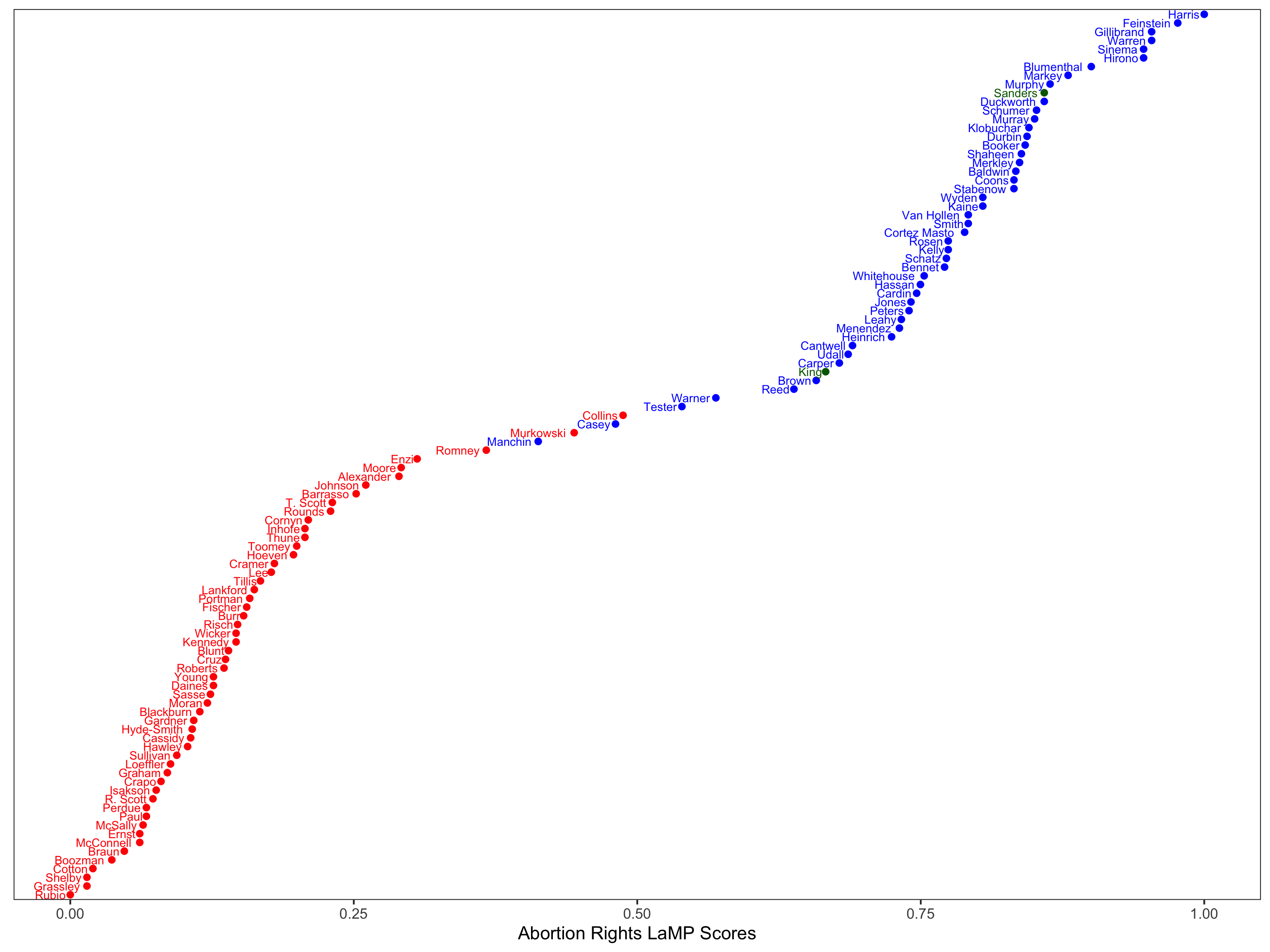}
    \caption{Abortion Rights LaMP scores across all senators. Democratic senators are in blue, Republican senators are in red, and Independent senators are in green.}
    \label{fig:overall_abortion_rights_lampscores_by_senator}
\end{figure}

The Abortion Rights LaMP scores of all senators of the 116th Congress are illustrated in Figure \ref{fig:overall_abortion_rights_lampscores_by_senator}. There is face validity with these scores. For example, it correctly separates the moderately pro-choice Republicans, Lisa Murkowski and Susan Collins. They are the only Republicans who describe themselves as pro-choice, although they often vote to confirm pro-life nominees. It also correctly separates Bob Casey and Joe Manchin, who self-describe themselves as pro-life and are endorsed by the Democrats for Life of America, a PAC that seeks to elect anti-abortion Democratic candidates. 

\subsubsection*{Abortion Rights LaMP scores better predict NARAL grades}
\begin{figure}[!ht]
    \centering
    \includegraphics[width=\textwidth]{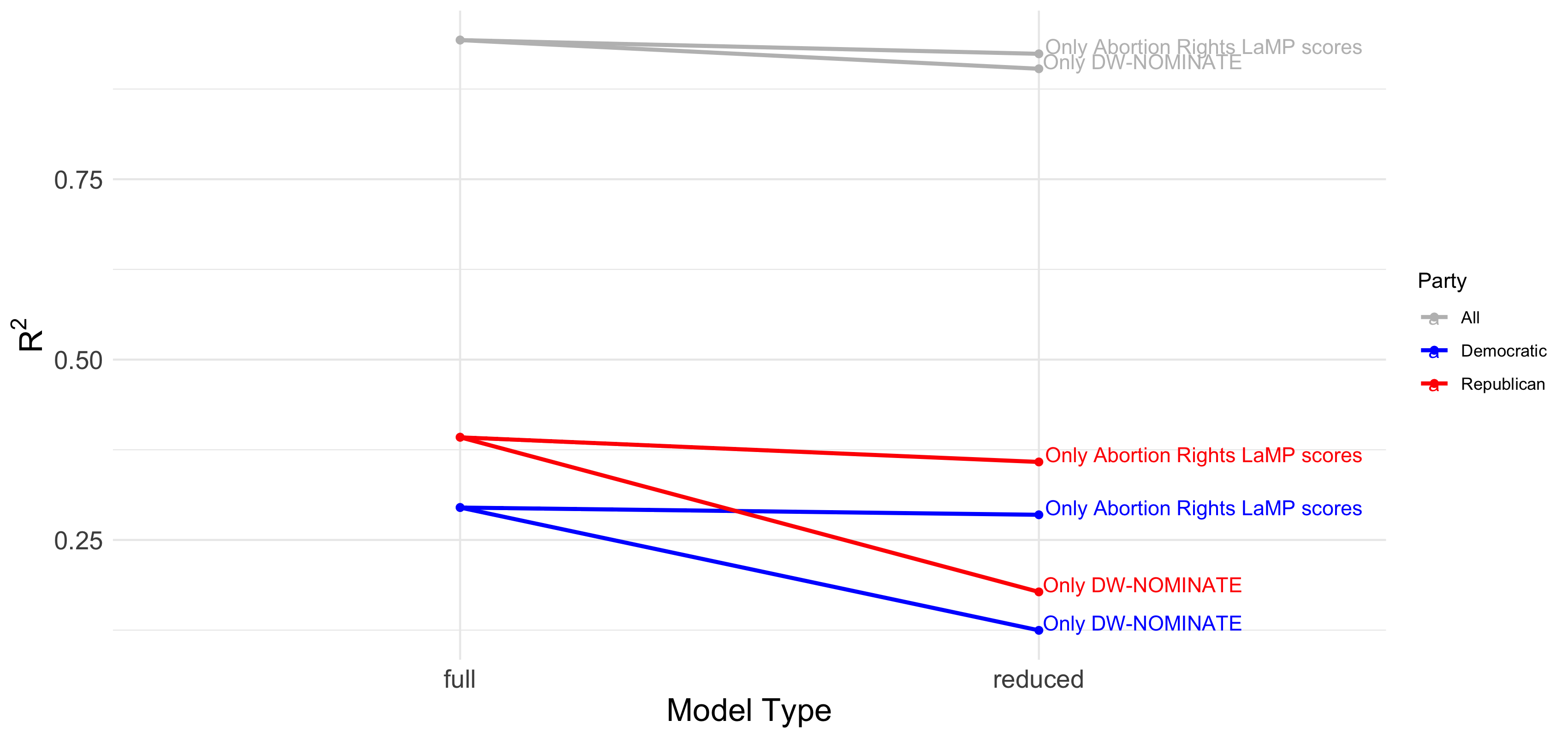}
    \caption{Comparing the proportion of variance explained, $R^2$, between the full and reduced models. The full model regresses the NARAL grades on both the first dimension of DW-NOMINATE and the Abortion Rights LaMP scores. The reduced models only use DW-NOMINATE or the Abortion Rights LaMP scores as predictors. The $R^2$ is always lower when Abortion Rights LaMP scores are removed as a predictor compared to when DW-NOMINATE is removed.}
    \label{fig:naral_predictive}
\end{figure}

We compare the predictive power of Abortion Rights LaMP scores and the first dimension of DW-NOMINATE on NARAL Pro-Choice America grades. NARAL assigns these grades using a set of votes on motions, bills, and confirmations that are related to abortion rights in some way. We used the NARAL grades from the end of 2020. We again use multivariate analyses to calculate how much the proportion of variance explained in NARAL grades falls when we compare the full model, regressing the NARAL grade on both the first dimension of DW-NOMINATE and Abortion Rights LaMP scores, with reduced models, which only use DW-NOMINATE or the Abortion Rights LaMP scores as the predictor. Figure \ref{fig:naral_predictive} shows how the $R^2$ drops as we move from full to reduced models across all senators, across the Democratic senators, and across the Republican senators.

We find that the proportion of variance explained in NARAL grades always falls more when we drop Abortion Rights LaMP scores as a predictor than when dropping DW-NOMINATE as a predictor. This holds for both Democratic and Republican senators. The partial F-test p-values for the full model and the reduced model with only DW-NOMINATE as a predictor are significant ($p < .001$) for both parties, while the p-value for the full model and the reduced model with only Abortion Rights LaMP scores are not ($p > .099$) for both parties. The difference in $R^2$ for models across all senators is negligible because stances on abortion still broadly fall along party lines. 

\section*{Discussion}
Our findings suggest that generative large language models can be useful for measuring the latent positions of lawmakers, especially on specific issues such as gun control and abortion. We find that the LLM is not hallucinating in these pairwise comparisons, LaMP scores are stable between repeated sets of matchups, and the LLM is not simply parroting existing scales such as DW-NOMINATE or interest group ratings. In other words, pairwise comparisons with an LLM yield sensible scales of lawmakers. Our evidence is consistent with the idea that the LLM synthesizes a great deal of information about lawmakers to evaluate latent constructs in predictable and sensible ways, agreeing with existing scales and predicting lawmaker behaviors like votes.

Our method is situated in a rapidly growing literature on using generative LLMs for social science applications. For example, these works have studied how generative LLMs can be used for labeling purposes \parencite{tornberg2023chatgpt4}, analyzing text along psychological constructs \parencite{rathje_mirea_sucholutsky_marjieh_robertson_vanbavel_2023}, reducing the divisiveness of online conversations \parencite{argyle2023ai}, and generating artificially politically extreme responses \parencite{bisbee_clinton_dorff_kenkel_larson_2023}. Most of these works focus on generating answers about one item at a time and studying how the LLM answers change across different items. On the other hand, our method examines how the LLM compares pairs of items and what kind of novel continuous measures can be derived using the LLM's answers. 

Our approach to scaling also speaks to a vast body of work on ideological scaling and ideal point estimation \parencite[see, e.g.,][]{keith_poole_nominate,poole1997ideology,heckman_snyder_1997,martin_quinn_2002,clinton_jackman_rivers_2004,wordfish_slapin_proksch,carroll_lewis_lo_poole_rosenthal,shor_mccarty_2011,lowe_benoit_2013,bonica2013,barbera_2015,temporao_vande_kerckhove_vanderlinden_dufresne_hendrickx_2018,wu_etal_2019,rheault_cochrane_2020,eady_bonneau_tucker_nagler_2020,hopkins_noel_2022,duck-mayr_montgomery_2023}. Estimation of ideology and stance has usually focused on behavior, such as how lawmakers vote in roll call votes or what specific words Twitter users use in tweets; alternative measures have focused on the perceptions of lawmakers, such as campaign donations, following-follower behavior on Twitter, and political activists' opinions. Our approach uses the embedded knowledge and analytical capabilities of LLMs, elicited through pairwise comparisons, about lawmakers' ideologies and their stances on public policy issues such as gun control and abortion to create continuous, unidimensional scales of these latent dimensions of politics and policy. 

Our approach has limitations due to the black box nature of deep neural networks: we do not know how ChatGPT is formulating its response to pairwise comparison prompts. We do know what texts were used to train ChatGPT-3.5 \parencite{gpt3_paper}---most notably Wikipedia and the Common Crawl, a corpus created from crawling the entire web---and we know that the existing measures discussed in this paper are present in the training data, as evidenced by its ability to define such measures correctly when prompted. One could attempt to ``understand'' why ChatGPT chooses a particular senator in each pairwise comparison using its outputted explanation. However, it may also use the answer as its context to generate the explanation. In other words, we still would not know what factors the LLM uses, why the LLM places greater weight on certain factors, nor would we know to what degree or what direction its explanation is related to its answer. The LLM cannot ``reason'' or ``deliberate'' about ideology or politics; it is, instead, a pattern recognition model. 

However, our analysis shows that the liberal-conservative ideology scale produced using an LLM's response to pairwise comparison prompts correlates well with multiple existing measures and reflects factors beyond position-taking. In other words, the findings above, such as the partial correlations and the Ideology LaMP scores' strong predictive power of activists' perceptions, suggest that LaMP scores reflect a blend of textual information: the scales are influenced by any number of factors that exist in its training corpora such as existing measures of ideology, floor speeches about certain public policy areas, news stories about senators, and people's expressed perceptions of these senators. LaMP scores can offset interpretive gaps in existing measures of ideology, such as dealing with lawmakers who vote against their own party for ideological reasons. In short, the LLM generates reasonable answers based on its training data.

What this means is that we can use LLMs with pairwise comparisons to estimate novel measures along specific political or policy dimensions, such as support for gun control or abortion rights, that could not be estimated using conventional scaling methods because of an absence of data related to behaviors or perceptions of the lawmakers. Estimating support for gun control among the same senators, we find that the scale not only differs in meaningful and intuitive ways from liberal-conservative scales but also that the scale \textit{predicts} votes on a gun control bill that is not in ChatGPT's training data. We find a similar pattern with the support of abortion rights. 

Despite some of its limitations, our proposed approach creates scales with both face validity and external validity. Our applications have been limited to American politics applications at the federal level, but one could extend the method to estimating the ideology of legislators in parliamentary systems where members of parliament vote strictly along party lines. A limitation of these comparisons, however, is that the LLM may not have enough information about every member of parliament. In such cases, a solution could be to pairwise compare text, such as campaign materials or tweets. Our approach can also be used to automatically evaluate whether the LLM's scaling of concepts of interest aligns with human judgments; this has potential applications for alignment research in natural language processing. In short, there are significant prospective future research contributions using our pairwise comparison approach for both the social sciences and natural language processing. 

\section*{Methods}
\subsection*{A Brief Overview of ChatGPT}
ChatGPT stands for Chat \textbf{G}enerative \textbf{P}retrained \textbf{T}ransformer. Given an initial text input called the prompt, ChatGPT generates a response. ChatGPT is built on GPT-3 \parencite{gpt3_paper}, a complex neural network (specifically, a decoder-only transformer) that predicts what token appears next given the set of existing tokens. It is trained on a massive corpus of text, which includes a filtered version of Common Crawl, WebText2, Books1, Books2, and Wikipedia. \textcite{jay_alammar_illustrated_gpt} explains and illustrates how decoder-only transformer large language models are trained in much greater detail. In short, its ability to produce coherent textual responses comes from the massive training corpora, the sheer size of the neural network (175 billion parameters), and the self-attention mechanism. The self-attention mechanism, in effect, allows the model to dynamically upweight and downweight certain parts of the input sequence \parencite{attention_is_all_you_need}. It achieves this by assigning a weight to each element of the input sequence that reflects the importance of that element relative to other elements.

ChatGPT is specifically trained to be a chatbot using a reinforcement learning technique called \textbf{r}einforcement \textbf{l}earning from \textbf{h}uman \textbf{f}eedback, or RLHF \parencite[see, e.g.,][]{early_rlhf}. RLHF works by first obtaining the generated text from the model. Human annotators then rank the output from most preferred to least preferred. A reinforcement learning algorithm---specifically, proximal policy optimization \parencite{ppo_algorithm}---is then used to update the large language model. For a more detailed discussion about RLHF, see \textcite{lambert2022illustrating}. The addition of RLHF training enables ChatGPT to generate human-like responses. It is important to note that ChatGPT can still produce incorrect responses (and sound extremely confident in doing so!), a phenomenon referred to as ``hallucination.'' 

It is also well-known that GPT-3 can generate text with biases, negative stereotypes, and unfair associations \parencite[see, e.g.,][]{lucy-bamman-2021-gender}. These biases can be potentially leveraged for social science purposes. For example, \textcite{argyle_busby_fulda_gubler_rytting_wingate_2023} find that biases in GPT-3 are fine-grained and demographically correlated and can be used to emulate partisan responses from a wide variety of human subgroups.

\subsection*{Using ChatGPT to Measure the Ideologies of Politicians}
\label{sec:chat_gpt_scaling}
We analyze the senators of the 116th Congress, which convened on January 3, 2019 and ended on January 3, 2021. We use this particular Congress because ChatGPT is trained on information up to 2021. We do not look at previous Congresses in order to prevent ChatGPT's newest information from leaking into assessments of the ideologies of members of previous Congresses. We obtain this list of senators from Voteview \parencite{voteview2021}. We keep Martha McSally (R-AZ) and Kelly Loeffler (R-GA) on the list of senators. Martha McSally was appointed to the Senate following interim Senator Jon Kyl's resignation. She then ran in Arizona's special election to finish the remainder of the Senate term but lost to Mark Kelly. Similarly, Kelly Loeffler was appointed to the Senate following Johnny Isakson's resignation for health reasons at the end of 2019.

We pairwise compare all senators of the 116th Congress. We call these pairwise comparisons ``matchups.'' We input the following prompt into ChatGPT for matchups between Democratic senators and matchups between a Democratic senator and a Republican senator: 
\begin{quote}
    \singlespacing
    Based on past voting records and statements, which senator is more liberal: [senator 1] ([senator 1 party abbrev]-[senator 1 state abbrev]) or [senator 2] ([senator 2 party abbrev]-[senator 2 state abbrev])?
\end{quote}
For matchups between Republican senators, we use a similar prompt:
\begin{quote}
    \singlespacing
    Based on past voting records and statements, which senator is more conservative: [senator 1] ([senator 1 party abbrev]-[senator 1 state abbrev]) or [senator 2] ([senator 2 party abbrev]-[senator 2 state abbrev])?
\end{quote}

We change the wording for matchups between Republican senators strictly because of a quirk of ChatGPT (and illustrative of its inability to ``reason'' about politics): when asked which senator in each pair is more liberal when comparing two conservative Republicans, it will often reply that neither senator is ``more liberal'' because they are both conservative Republicans. We use the default temperature parameter of 1 for ChatGPT-3.5. 

We use the ``more liberal'' prompt when comparing a Democratic senator and a Republican senator. Running pairwise comparisons using the ``more conservative'' prompt instead when comparing a Democratic senator and a Republican senator, the two Ideology LaMP scores correlate at 0.997. Using ``more liberal'' or ``more conservative'' for those particular matchups does not affect the LaMP scores. 

We record the name of the senator that ChatGPT considers to be more conservative (liberal) in each matchup. More specifically, we take ChatGPT's answer to the comparison prompt and use it in another prompt to extract the senator's name from the answer. See the Supplementary Information for more information about the prompt we used to extract the name of the more conservative senator from ChatGPT's response. Ties are allowed; this is when the LLM cannot assess who is more liberal (or more conservative) between the two senators. 

Between 102 senators, there are a total of 5,151 unique matchups. Each senator is compared to all other senators three times to study the consistency of ChatGPT's answers. The final comparison table contains each matchup with the number of wins (times a senator was deemed more conservative in that specific matchup) and losses (times a senator was deemed more liberal in that specific matchup) for that particular matchup. We consider ties 0.5 wins for both senators in the matchup; the section on the Bradley-Terry Model describes why we chose this approach for ties.

\subsection*{Using ChatGPT to Measure Gun Control Support}
To analyze the senators of the 116th Congress specifically for gun control support, we use the following prompt for all possible combinations of senators (5,151 comparisons).
\begin{quote}
    \singlespacing
    Based on past voting records and statements, who is more likely to support gun control: [senator 1] ([senator 1 party abbrev]-[senator 1 state abbrev]) or [senator 2] ([senator 2 party abbrev]-[senator 2 state abbrev])? 
\end{quote}
We then extract the names using a separate prompt; the Supplementary Information contains more information about the prompt we use to extract the name of the senator who is more likely to support gun control. Each senator is compared to all other senators three times to study the consistency of ChatGPT's answers; again, there are a total of 15,453 pairwise comparisons. We handle the wins, losses, and ties in the same way as the liberal-conservative ideology scale.

\subsection*{Using ChatGPT to Measure Abortion Rights Support}
To analyze the senators of the 116th Congress specifically for support of abortion rights, we use the following prompt for matchups between Democratic senators and matchups between a Democratic senator and a Republican senator. 
\begin{quote}
    \singlespacing
    Based on past voting records and statements, which senator is more pro-choice: [senator 1] ([senator 1 party-abbrev]-[senator 1 state abbrev]) or [senator 2] ([senator 2 party abbrev]-[senator 2 state abbrev])?
\end{quote}
For matchups between Republican senators, we used a similar prompt:
\begin{quote}
    \singlespacing
    Based on past voting records and statements, which senator is more pro-life: [senator 1] ([senator 1 party-abbrev]-[senator 1 state abbrev]) or [senator 2] ([senator 2 party abbrev]-[senator 2 state abbrev])?
\end{quote}
Again, we change the wording for matchups between Republican senators because ChatGPT will often not compare two Republican senators on a pro-choice comparisons basis. It will often only make a pairwise comparison when asked which Republican senator is more pro-life. 

\subsection*{Using the Bradley-Terry Model to Estimate Scales}
The Bradley-Terry model assumes that in a contest between two players $i$ and $j$, the odds that $i$ beats $j$ in a matchup are $\alpha_i / \alpha_j$, where $\alpha_i$ and $\alpha_j$ are positive-valued parameters that indicate latent ``ability'' \parencite{bradleyterry1952}. We can define $\alpha_i \equiv \exp(\lambda_i)$. Then, the log-odds of $i$ beating $j$ is 
\[\log\left[ \frac{\text{Pr}(i \text{ beats } j)}{\text{Pr}(j \text{ beats } i)} \right] = \lambda_i - \lambda_j\]
The intuition is that the larger the value of $\lambda_i$ compared to $\lambda_j$, the more likely it is for player $i$ to beat player $j$. 

We translate the above matchup into a contest between two senators regarding who is more conservative, more likely to support gun control, or more pro-choice. We use the liberal-conservative ideology scale as the running example in this section. The $\lambda$ parameters are measures of the senators' latent liberal-conservative ideology; the estimated parameters $\hat\lambda$ are the Ideology LaMP scores we describe in the Results section. We denote the more conservative senator in each matchup as the ``winner'' so that conservative senators have higher scores, intuitively matching the liberal-conservative political spectrum. For ties, we consider these 0.5 wins for both senators in the matchup. \textcite{bt_turner_firth} find that this approach yields ability parameter estimates that highly correlate with more complex approaches that explicitly deal with ties. We use the bias-reduced maximum likelihood estimation approach implemented in the \texttt{BradleyTerry2} R package with ChatGPT's responses to pairwise comparisons to estimate our scales of interest \parencite{bt_turner_firth}. The estimated parameters $\hat\lambda$ are relative to a reference senator (for all scales, we use Lisa Murkowski). However, this choice is unimportant because we rescale the estimated parameters to the unit interval to get the LaMP scores.

\newpage
\subsection*{Acknowledgements and Funding Sources}
We gratefully acknowledge that the Center for Social Media and Politics at New York University is supported by funding from the John S. and James L. Knight Foundation, the Charles Koch Foundation, Craig Newmark Philanthropies, the William and Flora Hewlett Foundation, the Siegel Family Endowment, and the Bill and Melinda Gates Foundation. We thank the members of the Center for Social Media and Politics for their helpful comments when workshopping this paper. We would also like to thank Maggie Macdonald and Megan Brown for their helpful comments throughout the paper-writing process.

\newpage 
\begin{singlespace}
\printbibliography[title={\large References}]
\end{singlespace}

\newpage
\appendix

\begin{center}
    \large
    Supplementary Information for ``Large Language Models Can Be Used to Estimate the Latent Positions of Politicians''\\
    \normalsize
    Patrick Y. Wu, Jonathan Nagler, Joshua A. Tucker, and Solomon Messing
\end{center}
\subsubsection*{Comparing Ideology LaMP Scores to the Second Dimension of DW-NOMINATE}
We also compare Ideology LaMP scores with the second dimension of DW-NOMINATE, which is typically interpreted as politicians' stance over cross-cutting social issues and sectional division \parencite{poole1997ideology,carroll_lewis_lo_poole_rosenthal}. In other words, the second dimension is interpreted as the stance of the major political parties over slavery, currency, nativism, civil rights, and lifestyle issues in various parts of American history. In recent Congresses, it has been interpreted as a measure capturing the anti-establishment and pro-establishment split \parencite{johnson_seconddim}. 

The correlation among all senators between Ideology LaMP scores and the second dimension of DW-NOMINATE is 0.388. This is approximately the same as the correlation between perceived ideology scores and the second dimension of DW-NOMINATE (0.412). It is also higher than the correlation between CFScores (0.317) and the first dimension of DW-NOMINATE (0.317) with the second dimension of DW-NOMINATE.

The correlation between LaMP scores and the second dimension of DW-NOMINATE is greater among Democratic senators (0.583) compared to Republican senators (0.027). If we interpret the second dimension of DW-NOMINATE to currently measure antiestablishmentarianism \parencite{johnson_seconddim}, it suggests that ChatGPT's responses correlate with the stances of Democratic senators towards the current party and government structures. On the other hand, ChatGPT's responses do not seem to correlate with anti-establishment or pro-establishment views of Republican senators.

\subsubsection*{Extracting the Name of the More Conservative Senator in Each Matchup}
For each pairwise comparison prompt, ChatGPT typically returns a small paragraph explaining its choice rather than returning only the name of the senator. For example, in a comparison between Ed Markey and Cory Booker, ChatGPT returned this answer: 
\begin{quote}
    \singlespacing
    Based on past voting records and statements, Ed Markey (D-MA) has generally been considered more liberal than Cory Booker (D-NJ).
    
    Ed Markey is known for his progressive stances on various issues such as climate change, renewable energy, and healthcare. He co-authored the Green New Deal resolution, a comprehensive plan to address climate change and create jobs. Markey has also been a vocal advocate for net neutrality, consumer protection, and stricter gun control laws. His voting record reflects consistent support for liberal causes and aligns with the Democratic Party's progressive wing. 
    
    Cory Booker, on the other hand, is often seen as a moderate within the Democratic Party. While he has supported liberal policies and causes, such as criminal justice reform and affordable housing, his stance has occasionally been more moderate than that of Ed Markey. Booker has emphasized issues related to economic inequality, racial justice, and expanding access to education. However, his positions on certain issues, such as healthcare, have been more centrist compared to some other progressive senators. It's important to note that political stances can evolve over time, and individual senators may take different positions on different issues. Therefore, it's always a good idea to refer to the most recent information and statements from the senators themselves to get the most accurate understanding of their current positions.
\end{quote}
We need to extract the names of the more liberal senators for Democratic senator matchups and Democratic-Republican senator matchups, and the more conservative senators for Republican senator matchups. To do this, we ask ChatGPT to extract the name. Specifically, we concatenate the above answer with the following prompt: 
\begin{quote}
    \singlespacing
    In the above Text, who is described to be the more liberal, more progressive, or less conservative senator: [senator 1] or [senator 2]? Return only the full name without party affiliation or state information. If one senator is described as more conservative, return the other senator's name. If one senator is described as more moderate, return the other senator's name. If neither senators are described to be more liberal, more progressive, less conservative, more conservative, or more moderate, reply with ``Tie.''
\end{quote}
For matchups where we prompt ChatGPT to return the name of the more conservative senator, we concatenate that answer with the following text:
\begin{quote}
    \singlespacing
    In the above Text, who is described to be the more conservative or less liberal senator: [senator 1] or [senator 2]? Return only the full name without party affiliation or state information. If one senator is described as more liberal, return the other senator's name. If one senator is described as more moderate, return the other senator's name. If neither senators are described to be more conservative, less liberal, more liberal, or more moderate, reply with ``Tie.''
\end{quote}
We then prompt ChatGPT with the concatenated text. ChatGPT usually returns the full name of the more liberal or more conservative senator. Punctuation and titles (such as appending ``Senator'' to the beginning of the name) were automatically removed using a Python function. Ties and answers that deviate from names are manually fixed. We also manually reviewed a sample of the answers that were given. There were occasional mistakes in the names extracted from the answers, but there did not seem to be a pattern in the mistakes. We also found that these mistakes were not repeated in repeated iterations of matchups. 

\subsubsection*{Extracting the Name of the Senator More Likely to Support Gun Control in Each Matchup}
Again, for each pairwise comparison prompt, ChatGPT typically returns a small paragraph explaining its choice rather than returning only the name of the senator. We concatenate the model's answer with the following prompt:
\begin{quote}
    \singlespacing
    In the above Text, which senator is described to be more likely to support gun control: [senator 1] ([senator 1 party abbrev]-[senator 1 state abbrev]) or [senator 2] ([senator 2 party abbrev]-[senator 2 state abbrev])? If one senator is described as being less likely to support gun control, return the name of the other senator. If one senator is described as more likely to support gun rights, return the name of the other senator. If neither senator is described to be more likely to support gun control, neither senator is described to be less likely to support gun rights, neither senator is less likely to support gun control, or neither senator is more likely to support gun rights, reply with ``Tie.'' Return only the full name without party affiliation or state information. Ignore any language about viewpoints changing.
\end{quote}
We then prompt ChatGPT with the concatenated text. Again, ties and answers that deviated from the names in the pairwise comparisons are manually fixed.

\subsubsection*{Extracting the Name of the More Pro-Choice Senator in Each Matchup}
We concatenate the model's answers with the following prompt for comparisons between Democratic senators or comparisons between a Democratic senator and a Republican senator in order to obtain the name of the senator who is more pro-choice in each matchup:
\begin{quote}
    \singlespacing
    In the above Text, which senator is described to be more pro-choice: [senator 1] ([senator 1 party abbrev]-[senator 1 state abbrev]) or [senator 2] ([senator 2 party abbrev]-[senator 2 state abbrev])? If one senator is described to be less pro-choice, return the name of the other senator. If one senator is described to be more pro-life, return the name of the other senator. Ignore any language about viewpoints changing. Return only the full name without party affiliation or state information. If both senators are described to be equally pro-choice, reply with ``Tie.''
\end{quote}
For matchups between Republican senators, the following prompt is used to extract the name of the senator who was more pro-life from the model's output:
\begin{quote}
    \singlespacing
    In the above Text, which senator is described to be more pro-life: [senator 1] ([senator 1 party abbrev]-[senator 1 state abbrev]) or [senator 2] ([senator 2 party abbrev]-[senator 2 state abbrev])? If one senator is described to be less pro-life, return the name of the other senator. If one senator is described to be more pro-choice, return the name of the other senator.  Ignore any language about viewpoints changing. Return only the full name without party affiliation or state information. If both senators are described to be equally pro-life, reply with ``Tie.''
\end{quote}

\end{document}